\documentclass[aps,prl,reprint,showpacs,superscriptaddress,longbibliography,nourl,noeprint]{revtex4-2}
\usepackage{amsmath,amssymb,physics,bm}
\usepackage{graphicx, microtype,siunitx} 
\usepackage[colorlinks=true,linkcolor=blue,citecolor=blue,urlcolor=blue]{hyperref}
\usepackage[all]{hypcap} 
\usepackage[colorinlistoftodos]{todonotes}

\newcommand{\rvec}[0]{{\bm r}} 
\newcommand{\Evec}[0]{{\bm E}} 
\newcommand{\nvec}[0]{{\bm n}}

\begin{document}


\title[]{Sculpting liquid crystal skyrmions with external flows}

\author{Rodrigo C. V. Coelho}
\affiliation{Centro de Física Teórica e Computacional, Faculdade de Ciências, Universidade de Lisboa, 1749-016 Lisboa, Portugal.}
 \affiliation{Departamento de Física, Faculdade de Ciências, Universidade de Lisboa, P-1749-016 Lisboa, Portugal.}
\email[]{rcvcoelho@fc.ul.pt}
\author{Hanqing Zhao}
\affiliation{Department of Physics and Soft Materials Research Center, University of Colorado Boulder, CO 80309, USA.}
\author{Mykola Tasinkevych}
\email{mykola.tasinkevych@ntu.ac.uk}
  \affiliation{Centro de Física Teórica e Computacional, Faculdade de Ciências, Universidade de Lisboa, 1749-016 Lisboa, Portugal.}
 \affiliation{Departamento de Física, Faculdade de Ciências,
Universidade de Lisboa, P-1749-016 Lisboa, Portugal.}
\affiliation{SOFT Group, School of Science and Technology, Nottingham Trent University, Clifton Lane, Nottingham NG11~8NS, United Kingdom.}
\affiliation{International Institute for Sustainability with Knotted Chiral Meta Matter, Hiroshima University, Higashihiroshima 739-8511, Japan.}
\author{Ivan I. Smalyukh}
\email{ivan.smalyukh@colorado.edu}
\affiliation{Department of Physics and Soft Materials Research Center, University of Colorado Boulder, CO 80309, USA.}
\affiliation{Department of Electrical, Computer, and Energy Engineering and Materials Science and Engineering Program, University of Colorado, Boulder, CO 80309.}
\affiliation{Renewable and Sustainable Energy Institute, National Renewable Energy Laboratory and University of Colorado, Boulder, CO 80309, USA.}
\affiliation{International Institute for Sustainability with Knotted Chiral Meta Matter, Hiroshima University, Higashihiroshima 739-8511, Japan.}
\author{Margarida M. Telo da Gama}%
  \affiliation{Centro de Física Teórica e Computacional, Faculdade de Ciências, Universidade de Lisboa, 1749-016 Lisboa, Portugal.}
 \affiliation{Departamento de Física, Faculdade de Ciências,
Universidade de Lisboa, P-1749-016 Lisboa, Portugal.}

\date{\today}

\begin{abstract}
We investigate, using experiments and numerical simulations, the distortions and the alignment of skyrmions in liquid crystal under external flows for a range of average flow velocities.
The simulations are based on the Landau-de Gennes $Q$ tensor theory both for isolated as well as for systems with many skyrmions. We found striking flow driven elongation of an isolated skyrmion and flow alignment of skyrmions in the many-skyrmion system, both of which are also observed in the experiments. In the simulations, particular attention was given to the dissipation rate and to the various dissipation channels for a single skyrmion under external flow. This analysis provides insight on the observed scaling regime of the elongation of isolated flowing skyrmions and revealed a surprising plastic response at very short times, which may be relevant in applications based on the alignment of soft structures such as liquid crystal skyrmions.

\end{abstract}

\maketitle



Liquid crystals (LCs) are unique soft materials that exhibit facile responses to external electric and magnetic fields. Additionally, the LC molecular orientation field, the director, can be easily distorted by anchoring at solid boundaries. These properties of LCs are the cornerstone of modern display technologies \cite{Chen2018ER}. In LC displays the director field undergoes continuous transitions between distinct topologically trivial configurations, i.e., configurations that can be morphed into a uniform structure continuously. Topological point or line defects may be spontaneously nucleated but these singular configurations are usually short-lived transient structures.  

Recently, a new type of topologically non-trivial non-singular director configurations was realised experimentally \cite{Smalyukh2010}. These solitonic structures are a soft matter analogue of magnetic skyrmions, which are two-dimensional solitons with localized winding of the magnetic moment observed in hard condensed matter systems \cite{muhlbauer2009,Hsu2017}, which are relevant for technological applications such as spintronics \cite{Krause2016}. Likewise, LC skyrmions are localized topologically protected distortions of the director field, which in some realisations can even include knotted or linked field lines \cite{manton_sutcliffe_2004,sutcliffe2017}.    
Additionally, skyrmions exhibit particle-like behaviour, such as long-ranged effective interactions between skyrmions and between skyrmions and colloids \cite{Sohn2019ER,sohn2018}, self-assembly into spatially periodic structures \cite{sohn2020}, and may acquire self-propelled motion when subject to periodic time-dependent electric fields  \cite{Ackerman2017,Zhao2023}.

As topological structures, skyrmions may be indexed or classified in terms of the elements of homotopy groups \cite{manton_sutcliffe_2004}. Experiments and numerical simulations revealed electric and magnetic transitions between skyrmionic textures with distinct topological indices \cite{tai2018}. It has been suggested that this behaviour may be used to control three-dimensional orientational structures of anisotropic nanoparticles, aligned with the local LC director, and drive them through transitions between different topological configurations \cite{zhang2015,Smalyukh_2020}. 

Despite the extensive body of research on LC skyrmions, their interaction with externally imposed material flow fields remains poorly understood. In our recent paper \cite{Coelho_2021} we reported a numerical study, based on the Ericksen-Leslie nematohydrodynamics, of the effect of external flows on the structures and dynamics of LC skyrmions. We found a configurational transition driven by the flow, from skyrmionic distortions along the flow in weak flows to skyrmionic distortions in the direction perpendicular to the flow in strong flows. 

Here, we extend that study and address explicitly the reversibility of the flow induced skyrmionic distortions. This analysis provides insight on the distortions observed in skyrmions under flow and reveals the mechanisms of their plastic response, which may be key in applications based on the elongation and alignment of soft structures such as liquid crystal skyrmions.

\begin{figure}[th]
\center
\includegraphics[width=0.8\linewidth]{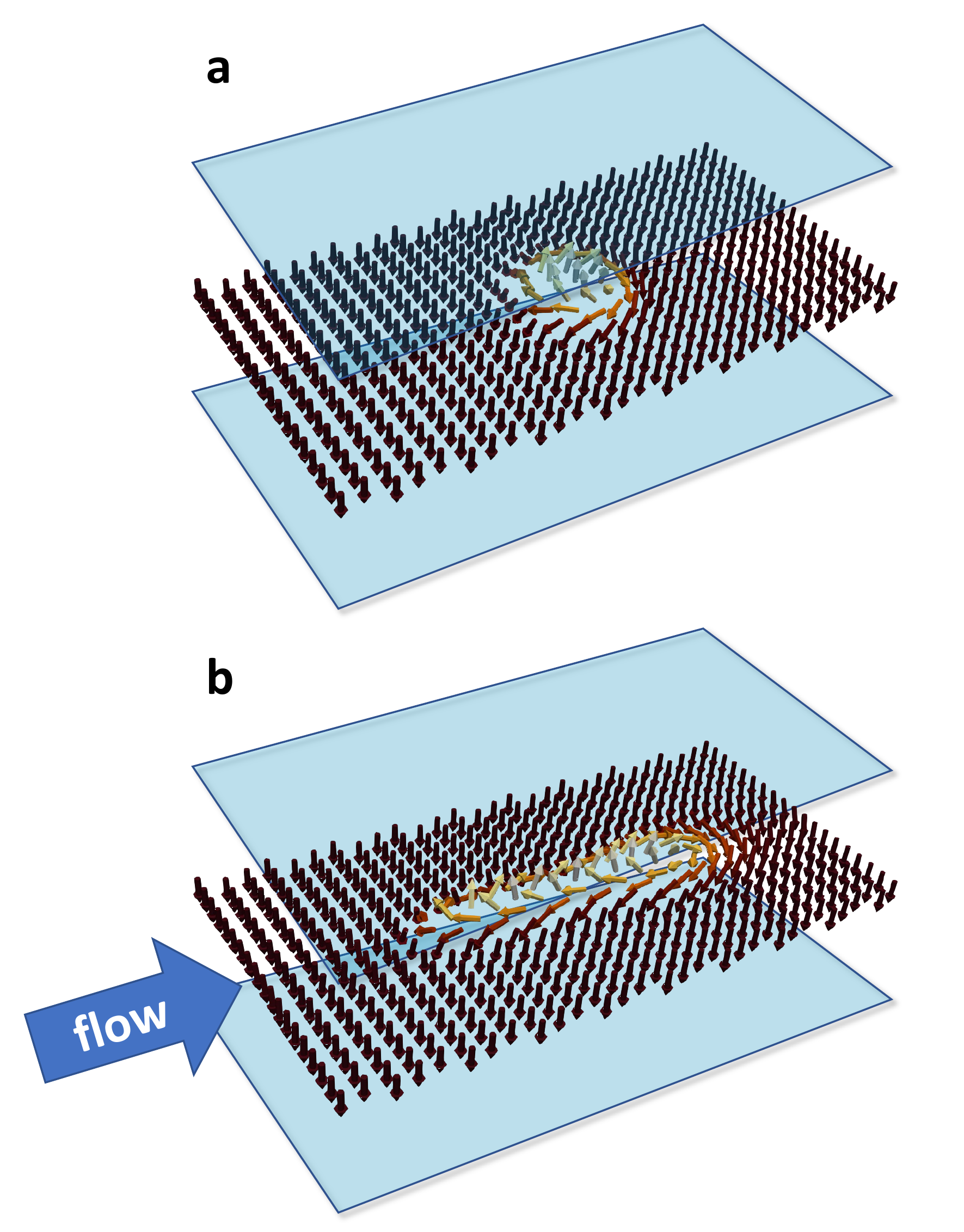}
\caption{{\bf Schematic illustration of the simulation setup}. a) Initial configuration of the director field. b) Elongated skyrmion as a result of the externally applied mass flow. The arrows represent the director field while the colours stand for its z-component (black for $n_z=-1$ and yellow for $n_z=1$). } 
\label{figure0-fig}
\end{figure}

We reformulate the problem in terms of the Landau-de Gennes $Q$ tensor theory and analyse the different contributions to the dissipated energy. Our numerical results for the flow-induced elongation of a skyrmion agree qualitatively with experimental observations. We also report on the collective behaviour of skyrmions driven by external flows observed both in the simulations and in experiments. The most significant many-body effects are the hindering of skyrmionic elongations by the presence of other skyrmions and the collective skyrmion alignment by the external flow. 
see Fig.~\ref{figure0-fig}

\begin{figure}[th]
\center
\includegraphics[width=0.75\linewidth]{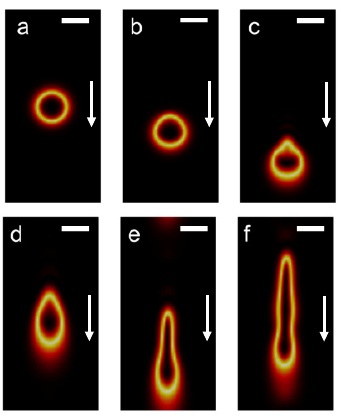}
\caption{{\bf Stretching of skyrmions by external flows.} Skyrmion configurations represented by colour coded $\vert n_z\vert$ at $\langle u \rangle =  4636.15\, \mu m/s$ (sample 2) at different times: (a) $t=0$, (b) $t=0.002 s$ ($t \approx t_{ch}$), (c) $t=0.02 s$ ($t \approx 10t_{ch}$), (d) $t=0.2 s$ ($t \approx 100t_{ch}$), (e) $t=0.6 s$ ($t \approx 300t_{ch}$), (f) $t=1 s$ ($t \approx 500t_{ch}$). Black corresponds to $\vert n_z\vert=1$ and yellow to $\vert n_z\vert =0$. The flow field is directed from the top to the bottom of the panels as is indicated by the white arrows. The scale bars correspond to one cholesteric pitch $p$.}
\label{figure1A-fig}
\end{figure}

\begin{figure}[th]
\center
\includegraphics[width=0.9\linewidth]{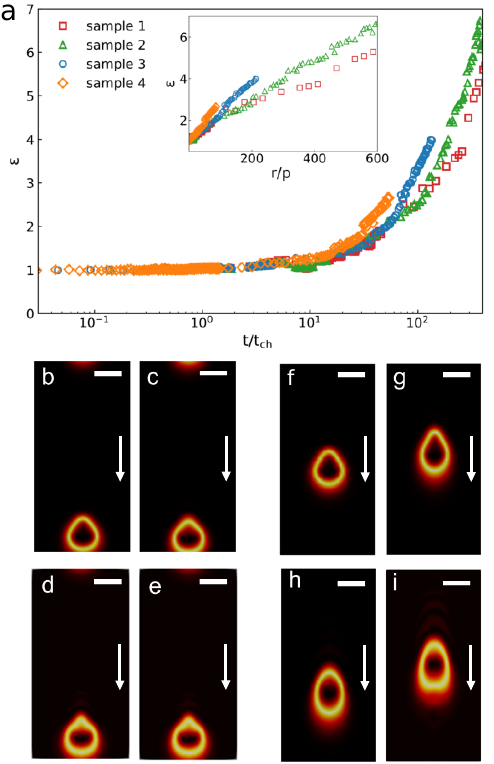}
\caption{ {\bf Dynamics of the skyrmion's elongation.} (a) Aspect ratio (length/width) of the flowing skyrmions as a function of the reduced time. Different symbols correspond to different values of the average flow velocity: sample 1 with $\langle u \rangle = 14671.36 \,\mu m/s$ ($t_{ch}\approx 0.0007 s$), sample 2 with $\langle u \rangle = 4636.15\, \mu m/s$ ($t_{ch}\approx 0.002 s$), sample 3 with $\langle u \rangle = 1467.14\, \mu m/s$ ($t_{ch}\approx 0.007 s$), and sample 4 with $\langle u \rangle = 463.62 \,\mu m/s$ ($t_{ch}\approx 0.02 s$). The time is scaled by the characteristic time $t_{ch}=p/\langle u \rangle$. The inset illustrates the aspect ratio as a function of the total displacement divided by the pitch. (b)-(e) Skyrmion configurations obtained at $t \approx 10 t_{ch}$ for different values of the average flow velocity:  (b) sample 4, (c) sample 3, (d) sample 2, (e) sample 1.  (f)-(i) Skyrmion configurations obtained at $t \approx 50 t_{ch}$ for (f) sample 4; (g) sample 3; (h) sample 2; (i) sample 1. The flow field is directed from the top to the bottom of the panels as is indicated by the white arrows. The scale bars correspond to one cholesteric pitch $p$.}
\label{figure1B-fig}
\end{figure}

\section{Results}
\label{sec:results}

\subsection{Flow-induced elongation of isolated skyrmions: numerics}
\label{one-skyrmion-sec}

We start by describing the elongation of a single skyrmion due to the externally imposed material flow. In the simulations, the flow is driven by an external constant force density field. The flow field rapidly reaches a uniform steady state, due to the friction at the surfaces, which enables an accurate description of the skyrmion behaviour under controlled conditions. 

Figure~\ref{figure1A-fig}(a)-(f) illustrates the time evolution of the shape of a single flow-driven skyrmion obtained at an average flow velocity $\langle u \rangle = 4636.15\, \mu m/s$ \textcolor{black}{($t_{ch}\approx 0.002 s$)}. Starting from a circular shape, Fig.~\ref{figure1A-fig}(a), the skyrmion becomes axisymmetric with a small tail-like wake, Fig.~\ref{figure1A-fig}(c), and undergoes continuous elongation in the flow direction (from the top to the bottom of the panel) beyond a characteristic time $t_{ch} = p/\langle u \rangle$, where $p$ is the cholesteric pitch. $t_{ch}$ is the time taken by the skyrmion to move by one cholesteric pitch, which is roughly the skyrmion size at $t=0$. The skyrmion elongation regime is illustrated in Fig.~\ref{figure1A-fig}(d)-(f), where the configurations are calculated at $t=100t_{ch}, 300t_{ch}, 500t_{ch}$, respectively.

In Fig.~\ref{figure1B-fig}(a), we plot the time evolution of the skyrmion aspect ratio or elongation $\varepsilon$, defined as the ratio of its length (parallel to the flow) and its width (perpendicular to the flow). We define the boundary of the skyrmion as the set of points where $\vert n_z \vert=0.5$. The four curves in Fig.~\ref{figure1B-fig}(a), corresponding to different flow velocities $\langle u \rangle$, collapse (approximately) onto a master curve when plotted against $t/t_{ch}$. The data collapse is better at short times $t\lesssim 10 t_{ch}$, which indicates that $t_{ch}$ is the relevant time scale in this regime. \textcolor{black}{The circular skyrmion shape starts to distort at $t\gtrsim 2t_{ch}$ (see Fig~\ref{figure1A-fig}(b), and configuration 2 in supplemental Figs.~S1-S4). Significant elongations, with $\varepsilon > 1.5$, are observed at $t\gtrsim 20 t_{ch}$ for the lowest $\langle u \rangle$ (orange diamonds in Fig.~\ref{figure1B-fig}(a)), and at $t\gtrsim 40 t_{ch}$ for the other velocities (blue circles, green triangles, and red squares in Fig.~\ref{figure1B-fig}(a)). At intermediate times of the order of $100 t_{ch}$ the skyrmion evolves into a commet-like shape as shown in Fig~\ref{figure1A-fig}(e), configuration 7 in Fig.~S1, and configuration 6 in Figs.~S2-S4, where $\varepsilon > 2$. Finally, at late times, we find, pearly string-like shapes as shown in Fig~\ref{figure1A-fig}(f), configuration 8 in Fig.~S1 and configuration 7 in Figs.~S2 and S3. This highlights the complex non-monotonic dynamics of the skyrmion distortions.}

We observe that at $t \gtrsim 30 t_{ch}$ when $\varepsilon \gtrsim 2.0$ (compare the orange diamonds with the other symbols in Fig.~\ref{figure1B-fig}(a)) the scaling of the elongation $\varepsilon$ starts to break down. In particular, the data for sample 4  (orange diamonds) shifts systematically upwards relative to the master curve, i.e. the  skyrmion elongation proceeds at a higher rate when $t \gtrsim 30t_{ch}$. Sample 3 follows the master curve until $t \approx 70 t_{ch}$ when $\varepsilon \approx 2.3$, and then shifts upwards similarly to the previous case. Finally, at $t \gtrsim 150 t_{ch}$, when $\varepsilon \approx 3.0$, the data for sample 2 (green triangles) departs from the data for sample 1. This cascade of thresholds where the scaling behaviour of the elongation breaks down can be related to skyrmion shape changes as illustrated in supplemental Fig.~S5(a)-(c), for samples 4-to-2 at times $t=30, 70$ and $150 t_{ch}$, respectively. All the configurations have a thin tail-like region on the left (Fig. S5(a)-(c)), which after forming lags behind the head-like front on the right, leading to effectively faster skyrmion elongations. For comparison, we also show in supplemental Fig.~S5(d) the configuration for sample 1 at $t=150t_ch$ which lacks a well defined tail-like feature. As a result this configuration elongates more slowly.

\begin{figure}[ht]
\center
\includegraphics[width=\linewidth]{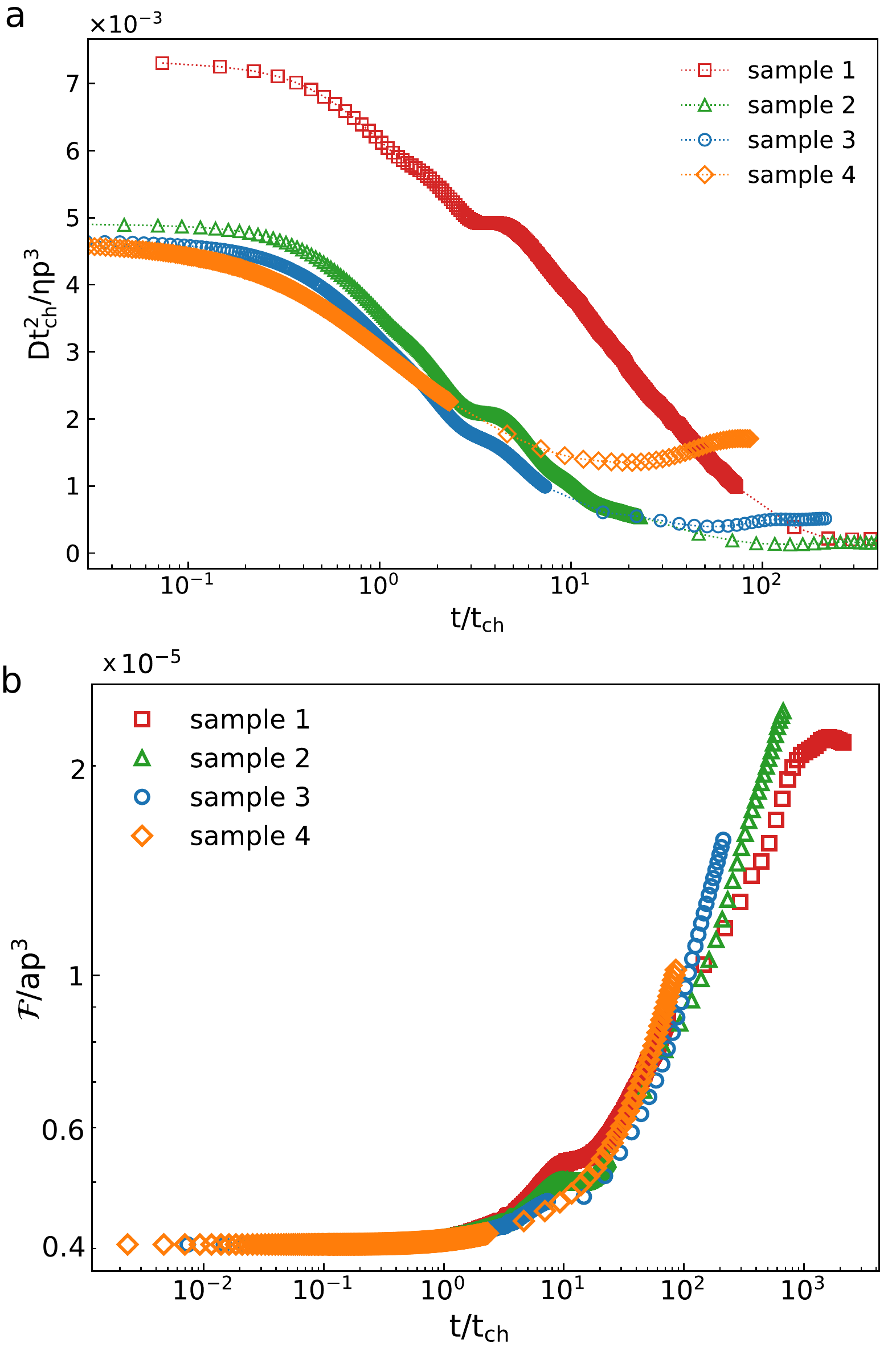}
\caption{{\bf Dissipation rate and Landau-de Gennes free energy.} (a) Dissipation rate $D$ in Eq.~\ref{dissipation-eq} as a function of time; (b) Landau-de Gennes free energy as a function of time, for the flow field switched on at $t=0$. Different symbols correspond to different values of the average flow velocity: sample 1 with $\langle u \rangle = 14671.36 \,\mu m/s$ ($t_{ch}\approx 0.0007 s$), sample 2 with $\langle u \rangle = 4636.15\, \mu m/s$ ($t_{ch}\approx 0.002 s$), sample 3 with $\langle u \rangle = 1467.14\, \mu m/s$ ($t_{ch}\approx 0.007 s$), and sample 4 with $\langle u \rangle = 463.62 \,\mu m/s$ ($t_{ch}\approx 0.02 s$).}
\label{dissipation-fig}
\end{figure}

\begin{figure}[ht]
\center
\includegraphics[width=\linewidth]{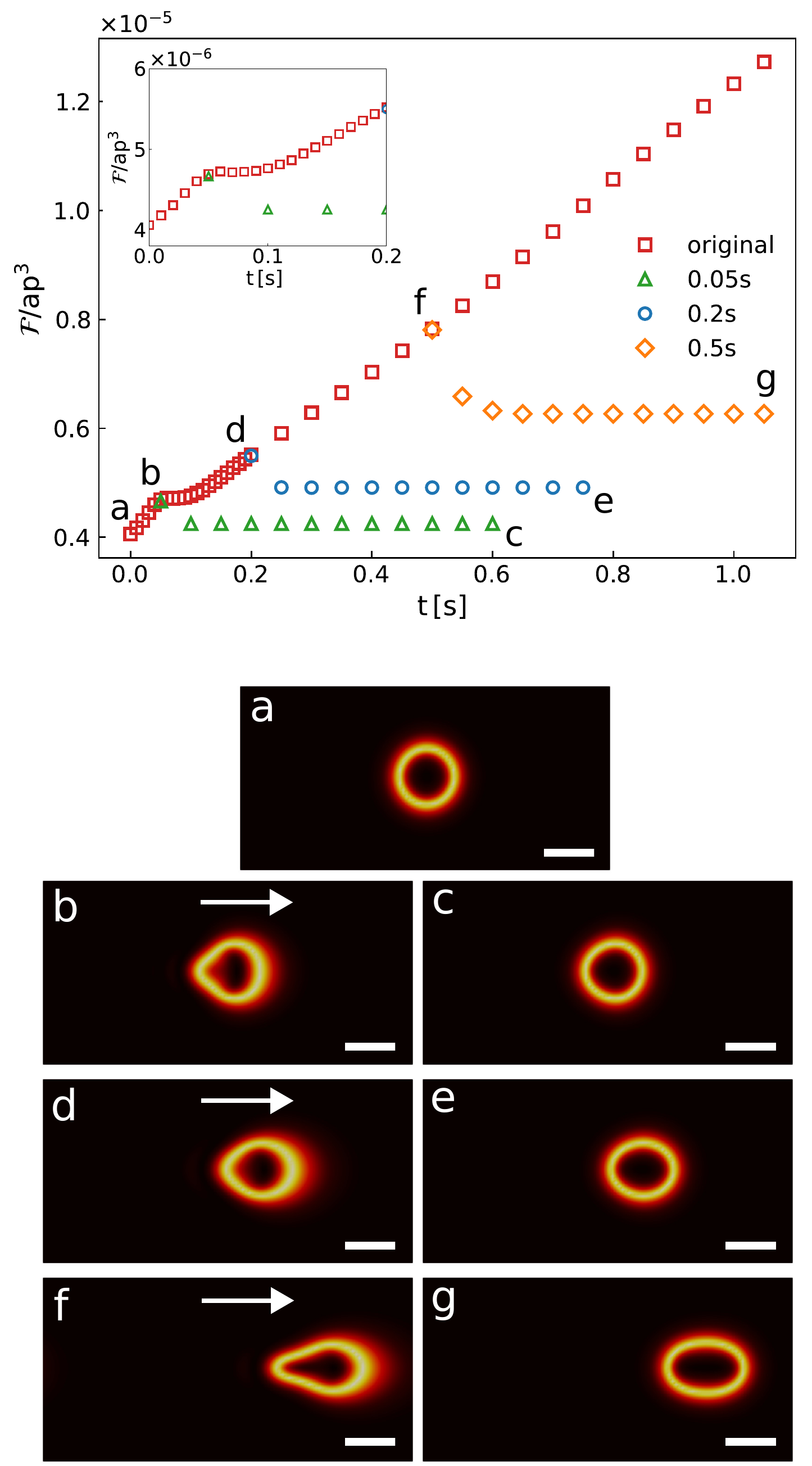}
\caption{{\bf Plastic nature of the skyrmion's elongation.} Top: Landau-de Gennes free energy as a function of time, for the flow field switched on at $t=0$ (red squares), and switched off at $t_{off}$. Triangles, circles and diamonds correspond to $t_{off}=0.05s (\approx 7t_{ch}), 0.2s (\approx 29 t_{ch})$ and $0.5s (\approx 71t_{ch})$, respectively. Data correspond to sample 3, with $\langle u \rangle = 1467.14\, \mu m/s$. The configuration on panel (a) is at $t=0$, and was obtained by minimizing the Landau-de Gennes free energy without flow. (b)-(g) correspond to the data labeled by the same letter on the top panel. The flow field is directed from the left to the right of the panels as is indicated by the white arrows. The scale bars correspond to one cholesteric pitch $p$.}
\label{free_energy-fig}
\end{figure}
\begin{figure*}[ht]
\center
\includegraphics[width=0.9\linewidth]{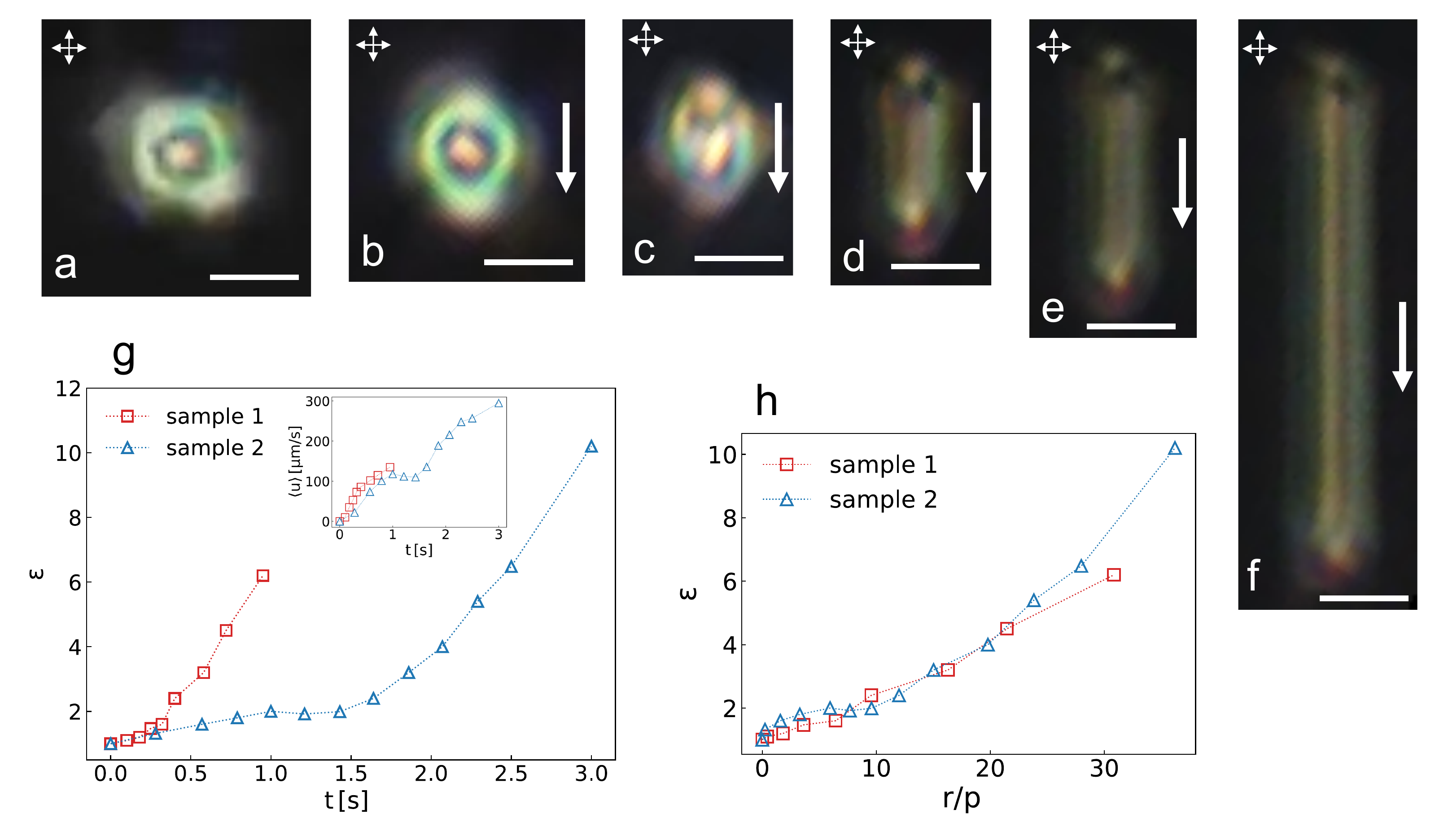}
\caption{{\bf Experimental realisation of a single flowing toron.} (a)-(g) micrographs of the toron obtained at times: (a) $0s$, (b) $0.18s$, (c) $0.32s$, (d) $0.4s$, (e) $0.58s$, (f) $0.95s$. The scale bar is $10\mu$m and the flow is from top to bottom as is indicated by the white arrows. (h) Aspect ratio and (inset) velocity against time for the two samples. (i) Aspect ratio against displacement divided by the pitch. Error bars in (h) and (i) are smaller than the size of the symbols.}
\label{figure1-2-fig}
\end{figure*}

With the aim of comparing the simulation results with experiments we depict in the inset of Fig.~\ref{figure1B-fig}(a) the aspect ratio $\varepsilon$ as a function of the skyrmion displacement $r$. The latter is defined as the distance between the skyrmion's centre of mass at a given time and its initial position. We recall, that the simulations stop for all samples at the same absolute time, corresponding to different reduced times and maximal skyrmion displacements. The curves exhibit a quasi-linear behaviour and good data collapse is observed for $r \lesssim 10 p$, in line with the scaling of $\varepsilon(t/t_{ch})$ discussed above. 

The linear behaviour is clear at low velocities $\langle u \rangle$, or large characteristic times (same pitch), see the orange diamonds in the inset of Fig.~\ref{figure1B-fig}(a). At higher velocities, we find $\varepsilon\approx 1 + 0.009 r/p$ for sample 2 (green triangles). The breakdown of scaling of the elongation at $r\gtrsim 10 p$ ($t\gtrsim 10 t_{ch}$), and the observation that in this regime $\varepsilon$ grows faster at lower $\langle u \rangle$ (compare the orange diamonds with the red squares in Fig.~\ref{figure1B-fig}(a)) are related to differences in the director field configurations at a given $t/t_{ch}$ for different flow velocities $\langle u \rangle$.

Figures.~\ref{figure1B-fig}(b)-(e) depict the skyrmion configurations under different flow velocities $\langle u \rangle$ at $t \approx 10t_{ch}$, while Figs.~\ref{figure1B-fig}(f)-(i) depict the configurations at a later time, $t\approx 50t_{ch}$. The first four configurations exhibit similar shapes and are in the scaling regime revealed in Fig.~\ref{figure1B-fig}(a). Significant shape changes, however, are observed at $t\approx 50 t_{ch}$ as illustrated by the last four configurations in Fig.~\ref{figure1B-fig}(f)-(i). Thus, the slowest skyrmion (Fig.~\ref{figure1B-fig}(f)) is already in the elongation regime, with well-pronounced head- and tail-like regions. By contrast, the fastest skyrmion (Fig.~\ref{figure1B-fig}(i)) has a bullet-like shape with a flat head (recall that the flow direction is from top to bottom.) Furthermore, the skyrmion shape becomes fuzzier as the flow velocity $\langle u \rangle$ increases. The differences between the samples are particularly clear when we analyse the flow field ${\mathbf u}({\mathbf r})$ shown in supplemental Fig.~S6(e)-(h). As the flow velocity $\langle u \rangle$ increases, the structure of ${\mathbf u}({\mathbf r})$ becomes irregular (see a wake-like feature in the rear of the skyrmion in supplemental Fig.~S6(h)), and the velocity fluctuations from the average ${\mathbf u}({\mathbf r}) - \langle u \rangle$  also increase.
\subsection{\textcolor{black}{Dissipation rate and Landau-de Gennes free energy}}
\label{egergies_numerics}

We proceed to relate the dynamics of the skyrmion distortions to the energy dissipation rate as a function of time. The work done by the external force density field (the term $\rho f_\alpha$ in Eq.~\ref{navier-stokes-eq}) is dissipated predominantly via the effective friction (the term $-\chi u_\alpha$ in Eq.~\ref{navier-stokes-eq}), which mimics the effect of the cell surfaces. The dissipation due to the friction has the form $\int_S d^2 r \chi \rho u^2$, and as we show below it varies little with time. Therefore, in what follows we discuss only the dissipation rate $D(t)$ arising from the liquid crystal flow and the director reorientation in the bulk. As we will show below, $D(t)$ sensitively probes the dynamics of the skyrmion distortions, and provides a better understanding of the failure of the scaling of the elongation $\varepsilon(t)$ at $t\geq 10t_{ch}$.

First, we note that the scalar order parameter $S$ hardly changes in the course of the simulations, and thus we use the Ericksen-Leslie theory to calculate the dissipated energy. We follow Ref.~\cite{marenduzzo07steady} to relate the parameters of the Beris-Edwards to those of the Ericksen-Leslie equations. In the Ericksen-Leslie theory the dissipated energy by a flowing liquid crystal is given by \cite{stewart2019static}:
\begin{align}
\label{dissipation-eq}
D &= \int_S d^2 r \Big[ \alpha_1(n_\alpha A_{\alpha\beta}n_\beta)^2 + 2\gamma_2 N_\alpha A_{\alpha\beta} n_\beta + \alpha_4 A_{\alpha\beta}A_{\alpha\beta}  \nonumber \\
&+(\alpha_5+\alpha_6)n_\alpha A_{\alpha\beta} A_{\beta\gamma} n_\gamma + \gamma_1 N_\alpha N_\alpha \Big],
\end{align}
where the $\alpha_n$'s are the Leslie viscosities, $\gamma_1=\alpha_3-\alpha_2$, $\gamma_2=\alpha_6-\alpha_5$ and the co-rotational time flux of the director is defined as $N_\alpha=\partial_t n_\alpha + u_\beta \partial_\beta n_\alpha - W_{\alpha\beta}n_\beta$. Diagonalisation of the $Q$ tensor field yields the director field up to a sign. Therefore, we carried out a standard procedure of director vectorisation. The resulting vector field is continuous and can be used to calculate the dissipated energy using Eq.~\ref{dissipation-eq}.  

We verified that upon applying the constant external force density, the fluid velocity away from the skyrmion relaxes almost instantaneously, as compared to the relaxation of the director field. Next, due to the 2D nature of this problem, this constant flow field away from the skyrmion cannot perturb the director field which has $n_z = 1$ in that region of the simulation domain. Therefore, the time dependence of all the contributions to the total dissipation rate discussed below is due solely to the director dynamics in the skyrmion proximal region.

We found that the most important contribution to $D(t)$ is due to the rotational viscosity term $\gamma_1 N_\alpha N_\alpha$ in Eq.~\ref{dissipation-eq}. Supplemental Fig.~S7, compares $D(t)$ with the dissipation rate $D_{1-5}(t)$ of the other five terms $\propto \alpha_1, \gamma_2, \alpha_4, (\alpha_5+\alpha_6)$ in Eq.~\ref{dissipation-eq}. The conclusion is that the contribution $D_{1-5}$ is an order of magnitude smaller than the dissipation arising from the $\gamma_1$ term. In the vicinity of the skyrmions the local flow differs from $\langle u \rangle$ (see Fig.~S6(e)-(h)), which results in a weak temporal variation of the frictional dissipation rate as shown in supplemental Fig.~S8. The quantity plotted is the reduced frictional dissipation rate $D_f(t)=\int_S d^2 r \chi \rho (u-\langle u \rangle)^2$, defined as the excess over the frictional dissipation for a uniform flow $\langle u \rangle$. $D_f(t)$ is two orders of magnitude smaller than $D(t)$, and thus can be safely ignored. 

Figure.~\ref{dissipation-fig}(a) illustrates $D(t)$ as a function of the reduced time $t/t_{ch}$ for several values of $\langle u \rangle$. We find that the three curves corresponding to the lower values of $\langle u \rangle$ collapse approximately onto a single curve, while the curve corresponding to the largest $\langle u \rangle$ stands apart. This can be understood qualitatively by inspecting the average velocity fluctuations $\Delta{\mathbf u}({\mathbf r})= ({\mathbf u}({\mathbf r})- \langle u \rangle)$, which is at least three times larger than for the other three samples (compare the colour bar in supplemental Fig.~S6(h) with those on Fig.~S6(e)-(g)). Large velocity fluctuations $|\Delta{\mathbf u}|$ enhance significantly the local velocity gradients, which in turn couple to the director field and drive it locally away from the preferred  orientation. This is one of the reasons why the dissipation rate for sample 1 (red squares in Fig.~\ref{dissipation-fig}(a)) lies markedly above the other three curves. \textcolor{black}{The regular symmetric flow pattern shows that the system is not in the turbulent regime.}

Despite this difference, the curves exhibit similar qualitative behaviour: 1) A short time linear decay $\approx (C -t)$, with $C$ a constant, for $t/t_{ch} \lesssim 1$. 2) An intermediate time effective power-law decay $\sim t^{\alpha(t)}$, for $1 \lesssim t/t_{ch} \lesssim t^*(\langle u \rangle)$, with estimated $-5/3 \lesssim \alpha(t) \lesssim -2/3$. 3) A late time quasi-constant regime, for $t \gtrsim t^*(\langle u \rangle)$, with $t^*/t_{ch}\approx 100, 20, 10, 2$ for samples 1 to 4, respectively. Interestingly, these values of the threshold time $t^*$ are an order of magnitude smaller than the characteristic time scale $\gamma_1 p^2 /L \approx 1.3 s$ (where $p$ is the cholesteric pitch and $L$ is the average elastic constant) for the relaxation of the director field.

\textcolor{black}{We emphasize that this separation into three regimes is only qualitative. For instance, we note a more complex behaviour of $D(t)$ within regime 2) for the three highest flow velocities, where we observed short plateau-like features for $3 \gtrsim t/t_{ch} \gtrsim 5$, see supplemental Figs.~S1-S3. The emergence of such features highlights the complex skyrmion distortion dynamics even when the skyrmion shape remains almost constant at short times, see the skyrmion configurations labelled 2 and 3 in supplemental Figs. S1-S3, for example. Interestingly, this stagnation of the skyrmion elongation is accompanied by a burst in the dissipation rate due to the coupling of the flow gradients to the director field as shown in the bottom panel of supplemental Fig.~S7. This panel reveals that the dynamical regime 1) $t/t_{ch} \lesssim 1$ is also observed in the relaxation of the dissipation term $D_{1-5}$.}      

\textcolor{black}{In regime 1) the dissipation rate $D(t)$ is reduced by a factor of two, while the skyrmion maintains its circular shape with $\varepsilon\approx 1$, as shown by the configuration depicted in Fig.~\ref{figure1A-fig}(b) obtained at the end of this regime, see also the configurations labelled 1 in supplemental Figs.~S1-S4. Next,  in regime 2) the total dissipation rate $D(t)$ is further reduced  by approximately an order of magnitude, while the elongation is $\varepsilon \lesssim 1.5$ for samples 2, 3 and 4 as shown in Fig.~\ref{figure1A-fig}(c) and supplemental Fig.~S2 (panel 5), Fig.~S3 (panel 4) and Fig.~S4 (panel 3). By contrast, for sample 1, we find a more significant skyrmion elongation at the end of regime 2), as shown in supplemental Fig.~S1 (panel 6).}

At late times, in regime 3) the skyrmions start to elongate and the total dissipation rate $D(t)$ takes an almost constant value (ca. 20 lower than the dissipation rate at early times). This behaviour is illustrated in supplemental Fig.~S1 (panels 7, 8), Fig.~S2 (panels 6, 7), Fig.~S3 (panels 6, 7), and Fig.~S4 (panels 5, 6). The elongation dynamics is more pronounced at the highest flow velocity (sample 1), panels 7 and 8 in supplemental Fig.~S1. The skyrmion configuration 8 extends almost throughout the whole system, reaching the limiting value of the aspect ratio. 
This fact is reflected by the saturation of the Landau-de Gennes free energy $F_{LdG}(t)$, as shown in Fig.~\ref{dissipation-fig}(b) (red squares for $t\gtrsim 1000 t_{ch}$). At shorter times, $F_{LdG}(t/t_{ch})$ exhibits a linear behaviour, and all the curves collapse when the time is scaled by the characteristic time $t_{ch}$.  A closer inspection of the curves, at the three highest flow velocities, reveals the existence of plateau-like behaviour at $10 \lesssim t/t_{ch} \lesssim 20$, where the skyrmion aspect ratio hardly changes, \textcolor{black}{as illustrated in supplemental Figs.~S1-S3 (panels 4 and 5). Again, this observation emphasizes a complex non-monotonic behaviour of the skyrmion elongation dynamics, which, among others, leads to the pearly string shapes shown in Fig~\ref{figure1A-fig}(f).}

We now proceed to discuss the reversibility of the skyrmion distortions induced by the flow. To this end, we calculate the Landau-de Gennes free energy as a function of time along two consecutive cycles when the flow field is switched on and off. Figure~\ref{free_energy-fig} illustrates the results for several values of the on-flow cycle. When the flow is on, after a short transient, the Landau-de Gennes free energy increases linearly with time (see open red squares in Fig.~\ref{free_energy-fig})--the work of the external force density is stored in the elastic distortions of the skyrmion. When the flow is switched off, a fraction of the stored elastic free energy is dissipated, and $F_{LdG}$ drops to a constant value, which depends on the duration of the preceding on flow cycle (see the open triangles, circles and diamonds in Fig.~\ref{free_energy-fig}). Somewhat unexpectedly, the free energy does not relax to its pre-flow value even when the flow was applied for periods as short as $7t_{ch}$, as shown by the green triangles in Fig.~\ref{free_energy-fig}. Correspondingly, the flow-elongated skyrmions (Figs.~\ref{free_energy-fig}(b), (d), and (f)) do not return to their original circular shape (Figs.~\ref{free_energy-fig}(a)) and remain in distorted metastable configurations (Figs.~\ref{free_energy-fig}(c), (e), and (g)). This reveals that flow-induced skyrmion distortions exhibit surprising plastic behavior even at elongations as small as $10\%$. A natural question arises whether one can design spatio-temporal flow patterns which would result in some pre-defined skyrmion shape?

\begin{figure*}[th]
\center
\includegraphics[width=0.75\linewidth]{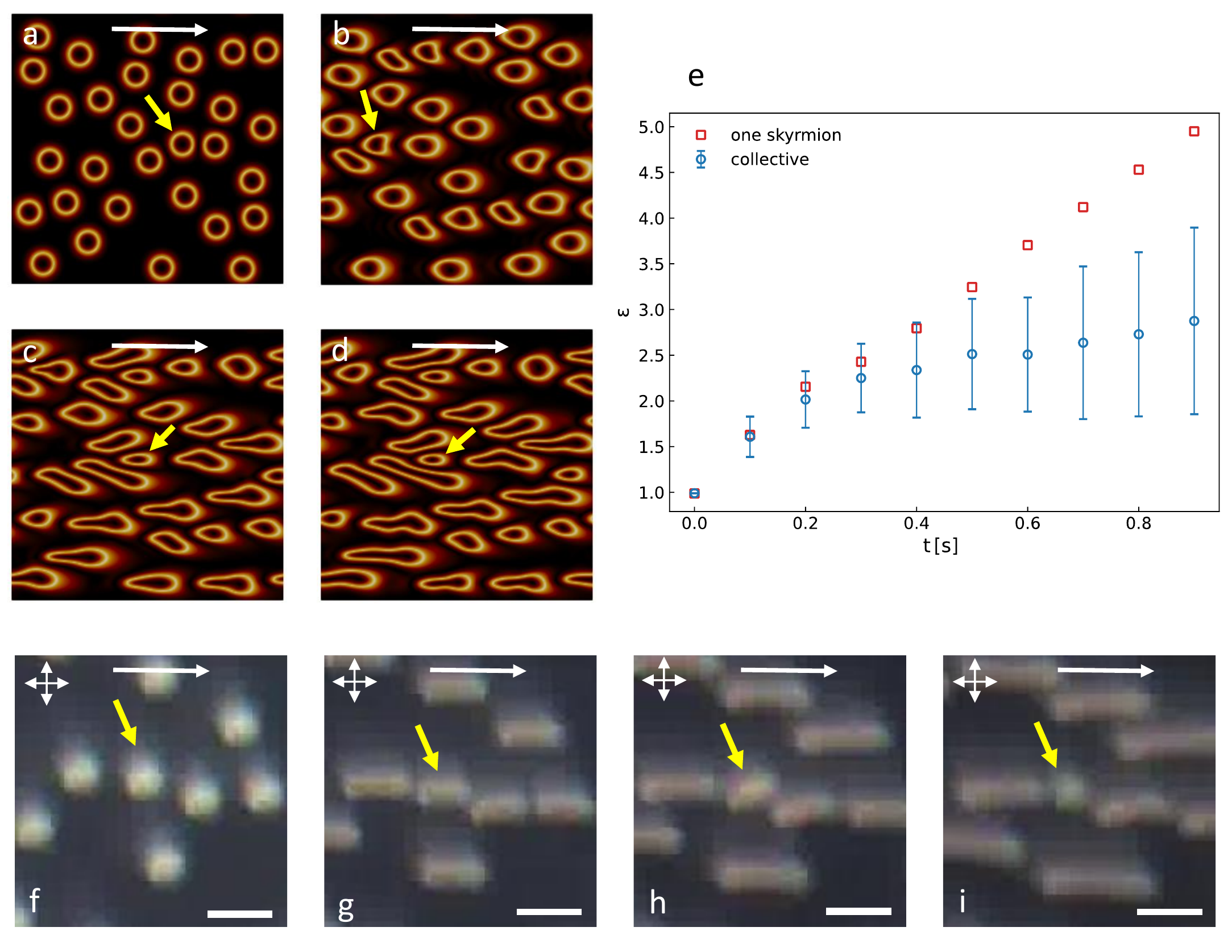}
\caption{{\bf Collective motion of flowing skyrmions.}
(a)-(d): Time evolution of the configurations of 30 skyrmions obtained numerically at $\langle u \rangle = 4636.15\, \mu m$. (a) $0$, (b) $0.1 s$, (c) $0.6 s$, (d) $0.9 s$. (e) Average aspect ratio as a function of time for the system with 30 flowing skyrmions depicted in (a)-(d). For comparison, the aspect ratio of a single skyrmion at the same flow velocity (sample 2) is also shown. The error bars represent the standard deviation of the skyrmions aspect ratio. (f)-(i): Time evolution of multiple torons in the experiments at (f) $t=0$, (g) $t=2s$, (h) $t=4s$ and (i) $t=6s$. The scale bar is $20\mu m$ and the flow is from left to right as is indicated by the white arrows. The yellow arrows indicate one structure (in the experiments and in the simulations) which does not elongate due to the interaction with its neighbours.}
\label{figure5-fig}
\end{figure*}

\begin{figure*}[th]
\center
\includegraphics[width=0.75\linewidth]{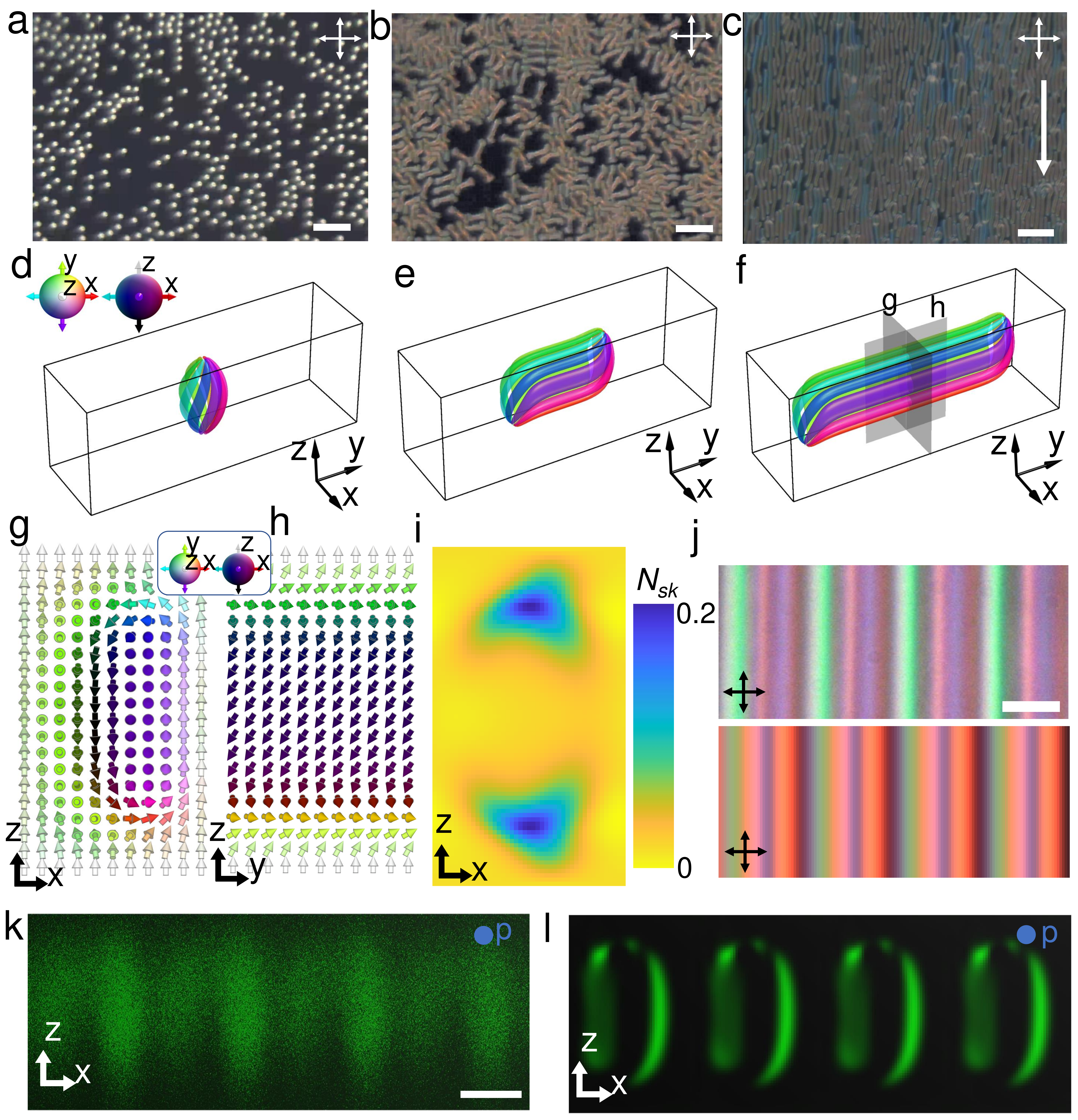}
\caption{{\bf Ordering of cholesteric fingers by the flow.} (a) and (b), time evolution of the cholesteric fingers without flow. In (a) the torons are stabilized by an applied sinusoidal voltage with the amplitude of $3.5V$ and the frequency $1000Hz$. In (b) the voltage is reduced and the torons destabilize forming cholesteric fingers of CF2 type \cite{sohn2018}.  The fingers elongate randomly until the entire space is filled. (c) time evolution of the CF-2 fingers with an applied flow. Initially, the torons are stabilized by the voltage as in (a), but then the voltage is reduced, and simultaneously an external flow is applied. This aligns the elongating torons along a common direction, and when the flow is turned off after $3s$ of operation, the fingers keep elongating in the same flow-imposed direction until they occupy the entire space (c). The scale bars in (a)-(c) are $50\mu$m. (d)-(f) Preimages show the schematic of a stretched toron of different sizes, where a preimage is the region in $\mathbb{R}^3$ which maps to a certain single point on the target order parameter space,$-$ real projective plane $\mathbb{RP}^2$. The colormap is shown in the inset of (d). (g) and (h) show detailed director field in  $(x,z)$ and $(y,z)$ cross-sections as marked by grey planes in (f). The director close to the cell surfaces is slightly tilted due to the flow drive. The colormap of different orientations of the director is shown in the inset. (i) The skyrmion number density $N_{sk}=\frac{1}{4\pi} \nvec\cdot \left (\frac{\partial\nvec}{\partial x}\times\frac{\partial\nvec}{\partial z}\right)$ of the director field in (g); $N_{sk}$ integrated over the whole cross-section in (g) equals unity. The colour scale for $N_{sk}$ is shown on the right. (j) Experimental (top) and computer-simulated (bottom) polarized optical micrographs of several aligned stretched torons. Light propagates along the $z-$direction. (k) Experimental and (l) computer-simulated three-photon excitation fluorescence polarizing microscopy images, the polarization of excitation light along the $y-$directions marked by {\em p}. Scale bars in (j) and (k) correspond to $10\mu$m.
}
\label{figure8-fig}
\end{figure*}

\subsection{Flow-induced elongation of isolated torons: experiments}
\label{one_skyrm_experiments}

It is challenging to control the average flow velocity in experiments due to the small size of the experimental cells. Injection of air into a cell with a syringe results in accelerated motion of the torons. Figures~\ref{figure1-2-fig}(a)-(g) are micrographs of toron configurations at different times as the flow proceeds through the cell. The toron elongation is similar to that observed in the simulations. Figure~\ref{figure1-2-fig}(h) depicts the aspect ratio as a function of time for two toron realizations and flow intensities, while the inset shows the average toron velocity as a function of time. As the toron velocity is not constant, the results plotted in Fig.~\ref{figure1-2-fig}(h) can not be compared directly with those in Fig.~\ref{figure1B-fig}(a). However, by plotting the aspect ratio as a function of the toron displacement, Fig.~\ref{figure1-2-fig}(i), we observe data collapse similar to that found in the simulations (inset of Fig.~\ref{figure1B-fig}(a)). This observation suggests that, if the toron velocities were constant, there would be a characteristic time for which the curves for the elongation $\varepsilon$ would collapse. It is noteworthy that even with the varying toron velocity, $\varepsilon$ is approximately linear in the displacement. For the data shown in Fig.~\ref{figure1-2-fig}(i) we find $\varepsilon\approx 1+0.17r/p$, where the slope is about 18 times larger than that found in simulations. We speculate that this difference may be due to the 2D approximation used in the simulations. Nevertheless, it is encouraging that even a 2D model agrees qualitatively with the experimental observations.

\subsection{Collective effects of flowing skyrmions/torons}

When many skyrmions are subject to a flow field, their elongation may be hindered or even reversed as a result of effective interactions mediated by distortions of the LC director. In Ref.~\cite{Coelho_2021}, two flowing skyrmions were simulated, and the resulting elongation was found to be smaller than that of an isolated skyrmion. The skyrmion-skyrmion interaction becomes relevant at distances of the order of the cholesteric pitch. 

We simulated the flow and distortions of 30 skyrmions initially placed at random on a square domain with dimensions $200\times 200$, for an average flow velocity $\langle u \rangle = 4636.15\, \mu m$ and periodic boundary conditions. Figures~\ref{figure5-fig}(a)-(d) illustrate the temporal evolution of the configuration of the ensemble of skyrmions. Most of the skyrmions elongate in the direction of the flow, with irregular transverse shape distortions due to skyrmion-skyrmion interactions. The elongation stops when the skyrmions fill the entire domain. We also observed that some skyrmions shrink as they are caged by neighbouring skyrmions, as shown by the arrows in Fig.~\ref{figure5-fig}(a)-(d). 

We calculated the average skyrmion aspect ratio and compared it with the single skyrmion case in Fig.~\ref{figure5-fig}(e). Initially, the two curves evolve similarly, but at $t\approx 0.2s$ they start to deviate indicating that a skyrmion in the ensemble elongates less than an isolated skyrmion. This occurs as the interactions with neighbouring skyrmions hinder the elongation of an isolated skyrmion. The error bars that increase with time indicate that the skyrmion distortions become more heterogeneous, as illustrated in Fig.~\ref{figure5-fig}(a)-(d). Similar behaviour was observed in experiments with ensembles of flowing torons as shown in Fig.~\ref{figure5-fig}(f)-(i). The toron indicated by the arrow initially elongates and then shrinks back to an almost circular shape as the result of interactions with the elongating neighbours. 

Clearly, the elongation of flowing skyrmions discussed in the previous section occurs only if they are isolated or if their elongation proceeds along a common direction. The time dependence of the free energy of the system with 30 skyrmions is similar to that of the average aspect ratio (see supplemental Fig.~S9). At short times it is similar to that of an isolated skyrmion, flowing at the same average velocity, and it exhibits many-body effects beyond $t\approx 0.2$s as the interactions between the elongated skyrmions become relevant. At late times the free energy increases more slowly in line with the suppression of the average elongation of the skyrmions.

\section{Discussion}

We investigated, using experiments and numerical simulations, the distortions and the alignment of liquid crystal skyrmions under external flows for a range of average flow velocities.

We performed extensive simulations based on the Landau-de Gennes $Q$ tensor theory both for isolated as well as for systems with many skyrmions. The most striking effects were the flow driven elongation of single skyrmions and their alignment along the flow direction in the many skyrmion system, both of which were also observed in the experiments. 

The simulations were run for different flow velocities $\langle u \rangle$ and the results revealed a characteristic time scale $t_{ch} = p/ \langle u \rangle$, which controls the elongation of single skyrmions, leading to data collapse or scaling of $\varepsilon$ as a function of $t/t_{ch}$.

Specific features were observed depending on $\langle u \rangle$: the larger the flow velocity $\langle u \rangle$, the larger the extention of the scaling regime. This was related to a configurational transformation where the skyrmion takes a comet-like shape where the tail moves more slowly than the head. In other words the skyrmion is pulled apart by the flow, with distinct regions responding to the flow in different manners.

Furthermore, due to finite size effects, observed at high flow velocities and long times, the free energy flattens or saturates and remains almost static as the flow induced elongation is hindered by the image skyrmion that results from the periodic boundary conditions. 

We have also calculated the dissipation rate and the various dissipation channels for a single skyrmion under external flows. This analysis provided insight on the observed scaling regime of the elongation of a single skyrmion and revealed a somewhat surprising plastic response at very short times. The plastic character of the skyrmion elongation occurs even at small strains, where one would expect an elastic response to the forcing. This plastic behaviour is expected to be present also in the many skyrmion system and may be relevant in applications based on the alignment of soft structures such as liquid crystal skyrmions.

The experimental observations of the flow induced distortions and alignment of single and many-skyrmion systems agree qualitatively with the numerical results, despite important differences between the model and the real system. We single out the fact that the simulations were carried out in 2D under controlled uniform flow conditions, while the experiments are in 3D and the flow was not uniform. Nevertheless the elongation and the alignment of the distorted skyrmions by external flows appear to be robust and are qualitatively similar. The plastic behaviour uncovered by our analysis of the dissipation may also have an impact on experimental observations and applications.

We finish by reporting an interesting experimental observation that is not reproduced by the 2D model. Under certain experimental conditions, skyrmions have been observed to de-stabilize and transform into cholesteric fingers, which elongate in random directions. Preliminary work suggests that applying a flow field for a short time results in orientational ordering of the elongated cholesteric fingers, which persists when the flow is switched off. 

Previous experimental work reported that torons were stabilized by applying a weak voltage across the cell and were observed to transform into cholesteric fingers~\cite{PhysRevE.72.061707} when reducing this voltage. The fingers grow in random directions triggered by a decrease of the voltage below a certain threshold. Each toron (initially with a spherical shape in Fig.~\ref{figure8-fig}(a)) elongates in a random direction (Fig.~\ref{figure8-fig}(b)) and continues to grow until filling the available space. 
Figures~\ref{figure8-fig}(c) shows growing fingers, but now under an applied external flow that is switched on for a short time just after the voltage reduction. We found that it was sufficient to maintain the flow for 3$s$ in order to align all the fingers in the flow direction, as shown in Fig.~\ref{figure8-fig}(c). Moreover, after switching the flow off, the fingers continue to elongate along the same direction, filling the entire space. This suggests that an external flow may be effective to drive orientational ordering of cholesteric fingers. Interestingly, this ordered texture of fingers remains stable for days, although, some of the fingers shrink back to their original circular shape under an effective compressing action of the neighbouring fingers. 

\textcolor{black}{There exist several distinct types of cholesteric fingers \cite{sohn2018}. To determine the type of the fingers observed here, we carried out a quasi-static numerical analysis of the 3D structure of the elongated torons, by minimizing the Frank-Oseen elastic free energy. Several, examples of the stretched toron configuration are presented in Figs.~\ref{figure8-fig}(d)-(f), using colour coded director preimages, i.e., those spatial regions where the nematic director $\nvec$ has a given fixed value. The vectorial representations of $\nvec(\rvec)$, shown in Fig.~\ref{figure8-fig}(g) and (h), reveals that the topology of this extended structure remains unchanged, because the vectorized $\nvec(\rvec)$ taken on the $(x,z)$ cross section of this extended configuration covers the order parameter space $\mathbb{S}^2$ once. This finding is also confirmed by directly verifying that the skyrmion number $\frac{1}{4\pi} \int \nvec\cdot \left (\frac{\partial\nvec}{\partial x}\times\frac{\partial\nvec}{\partial z}\right) dx dz = 1$. The corresponding skyrmion number surface density $N_{sk}$ in Fig.~\ref{figure8-fig}(i) highlights the existence of two nonsingular $\lambda-$disclinations, or fractional skyrmions \cite{Wu:2021},  at the top and the bottom of the cross section. In fact, this configuration is characteristic of a cholesteric finger of the second type, CF-2, studied previously \cite{PhysRevE.57.3038}, and is often found as a metastable configuration occurring spontaneously. The director configuration in CF-2 shows how the fractional values of the skyrmion number, two half skyrmions in this case, can add up to unity in the finger, being embedded into the uniform far field background.}

\textcolor{black}{Figures \ref{figure8-fig}(j)-(l) compares computer-simulated and experimental polarized optical microscopy (POM) as well as three-photon excitation fluorescence polarizing microscopy ((3PEF-PM) images for a system of aligned extended torons. We observe very good qualitative agreement, which demonstrates that the experimental elongated structures are cholesteric fingers of the second type. It is natural to assume that similar director configurations are associated with the torons stretched by the flow, being subject to stabilising voltage at the same time. More detailed numerical investigation of these 3D structures and their interaction with the external flow is beyond the scope of this study, which will be addressed in a future work.}

Other extensions of this study include the effect of a time-dependent, e.g. step-like, flow field, the  introduction of local changes in the flow direction or the effect of flows with non-zero vorticity.

\label{conclusion-sec}

\section{Methods}
\label{method-sec}

\subsection{Beris-Edwards model for flowing skyrmions}

The model used to describe the liquid crystal skyrmions under mass flow is based on the Landau-de Gennes $Q$ tensor theory of nematics~\cite{p1995physics, Doi2013}. For uniaxial ordered phases, the tensor order parameter is $Q_{\alpha \beta} = S(n_\alpha n_\beta - \delta_{\alpha \beta}/3)$, where $S$ is the scalar order parameter, which equals to zero in the isotropic phase and unity in phases with perfect alignment, and $n_\alpha$ is the director field. The free energy of the liquid crystal is $\mathcal{F} = \int_V \,d^3 r\, f$, where the free energy density $f$ has three contributions:
\begin{align}
 f_b=& a \, {Q_{\alpha \beta}}^2 - b \, (Q_{\alpha \beta} Q_{\beta \gamma} Q_{\gamma \alpha})   +c \, {Q_{\alpha \beta}}  ^4, \\ 
 f_{el}=& \frac{L_1}{2} (\partial _\gamma Q_{\alpha \beta})^2 + \frac{L_2}{2} \partial _\epsilon Q_{\nu\epsilon} \partial_{\gamma} Q_{\nu \gamma} \nonumber \\
& + \frac{4\pi L_1}{p} \varepsilon_{\alpha\beta\gamma} Q_{\alpha\epsilon}\partial_\beta Q_{\gamma \epsilon} , \\
 f_w =& \frac{W}{2}\left( Q_{\alpha\beta} - Q^0_{\alpha\beta} \right)^2. 
\end{align}
The first is the bulk free energy and $a$, $b$ and $c$ are positive material constants. The second is the elastic free energy with two elastic constants $L_1$ and $L_2$~\cite{PhysRevE.94.062703,doi:10.1098/rsta.2020.0394} and $p$ is the equilibrium cholesteric pitch. The last contribution arises from the anchoring at the confining surfaces. As the simulations are effectively 2D, one needs to apply this anchoring everywhere (as a bulk term) to stabilize the skyrmions~\cite{Coelho_2021,C9SM02312G,PhysRevE.101.042702}. The parameter $W$ controls the anchoring strength.

The dynamics is governed by the Beris-Edwards, the continuity and the Navier-Stokes~\cite{beris1994thermodynamics} equations:
\begin{align}
  &\partial _t Q_{\alpha \beta} + u _\gamma \partial _\gamma Q_{\alpha \beta} - S_{\alpha \beta} = \Gamma H_{\alpha\beta} , \label{beris-edwards-eq} \\
&  \partial _\beta u_\beta = 0, \label{continuity-eq}\\
&\rho\partial_t  u_\alpha + \rho u_\beta \partial _\beta   u_\alpha  = -\chi u_\alpha + \partial_\beta [ 2\eta A_{\alpha\beta}  + \sigma^{\text{n}}_{\alpha\beta} +\rho f_\alpha],   \label{navier-stokes-eq}
\end{align}
where $\rho$ is the liquid crystal density and $\Gamma$ is the system dependent rotational diffusivity.
Equation~\ref{beris-edwards-eq} describes the time evolution of the order parameter $Q_{\alpha\beta}$, which depends on the velocity field $u_\alpha$. The dynamics of the velocity field is given by Eq.~\ref{navier-stokes-eq}, which depends on $Q_{\alpha\beta}$. The first term on the right-hand side of Eq.~\ref{navier-stokes-eq} stands for the effective friction at the surfaces~\cite{Coelho_2021}. For Poiseuille flow of a fluid with absolute viscosity $\eta=\rho\nu$ between two surfaces separated by a distance $L$, the friction coefficient in 2D that yields the same average velocity of the 3D flow is $\chi=12\eta/L^2$, where $\nu$ is the kinematic viscousity. The fluid moves driven by an external force $f_\alpha$. In the equations above, the shear rate and the vorticity are given by $A_{\alpha\beta} = (\partial_\alpha u_\beta + \partial_\beta u_\alpha)/2$ and  $W_{\alpha\beta}= (\partial _\beta u_\alpha - \partial _\alpha u_\beta )/2$ and  the co-rotational term reads:
\begin{align}
 &S_{\alpha \beta} = ( \xi A_{\alpha \gamma} + W_{\alpha \gamma})\left(Q_{\beta\gamma} + \frac{\delta_{\beta\gamma}}{3} \right) \nonumber \\
&
 + \left( Q_{\alpha\gamma}+\frac{\delta_{\alpha\gamma}}{3} \right)(\xi A_{\gamma\beta}-W_{\gamma\beta}) 
 \nonumber \\
&
 - 2\xi\left( Q_{\alpha\beta}+\frac{\delta_{\alpha\beta}}{3}  \right)(Q_{\gamma\epsilon} \partial _\gamma u_\epsilon), 
\end{align}
where $\xi$ is the flow aligning parameter. The molecular field is: 
\begin{align}
 H_{\alpha\beta} = -\frac{\delta \mathcal{F}}{\delta Q_{\alpha\beta}} + \frac{\delta_{\alpha\beta}}{3} \Tr \left( \frac{\delta \mathcal{F}}{\delta Q_{\gamma \epsilon}} \right)
\end{align}
and the nematic stress tensor is given by:
\begin{align} 
 \sigma_{\alpha\beta}^{\text{n}} &= -P_0 \delta_{\alpha\beta} + 2\xi \left( Q_{\alpha\beta} +\frac{\delta_{\alpha\beta}}{3} \right)Q_{\gamma\epsilon}H_{\gamma\epsilon}  \nonumber \\ 
 & - \xi H_{\alpha\gamma} \left( Q_{\gamma\beta}+\frac{\delta_{\gamma\beta}}{3} \right) - \xi \left( Q_{\alpha\gamma} +\frac{\delta_{\alpha\gamma}}{3} \right) H_{\gamma \beta}   \nonumber \\ 
 & - \frac{\delta \mathcal{F}}{\delta (\partial_\beta Q_{\gamma\nu})}\, \partial _\alpha Q_{\gamma\nu}  + Q_{\alpha\gamma}H_{\gamma\beta} - H_{\alpha\gamma}Q_{\gamma\beta} ,
\end{align}
with $P_0$ the hydrostatic pressure.

\subsection{Numerics}

As discussed in Ref.~\cite{Coelho_2021}, the time scales for changes in the director and the velocity fields differ by six orders of magnitude, rendering the simulations challenging. The simulations reported in  Ref.~\cite{Coelho_2021} use finite differences for both fields ($n_\alpha$ and $u_\alpha$) and an adaptive time step to speed up the convergence of the velocity field. The dynamics of the director field was described by the Ericksen-Leslie theory. It was also assumed that the fluid relaxes instantaneously when compared to the director field. Here, we use the Beris-Edwards equation for the dynamics of the director field, which is solved using finite differences (predictor-corrector algorithm) while the lattice Boltzmann method is used for the velocity field. This approach results in slower but more reliable simulations, since the dynamics of the two fields are solved with the same time step. In addition, the lattice Boltzmann method is conservative while the method based on finite differences is not. Perhaps the major advantage of this approach is that the simulations are stable for parameters close to the experimental ones, which was not the case for the Ericksen-Leslie theory~\cite{Coelho_2021}.

Due to the different time scales, which require long simulations, a compromise between the domain size and the simulation time is required. We used 2D simulation boxes, which provide useful insights into the dynamics of skyrmions in feasible simulation times. However, a 2D domain assumes invariance in the direction perpendicular to the plane, which may be a strong approximation for the velocity field. 
In addition, the skyrmion structure in the experiments is not invariant in the direction perpendicular to the plane. Thus, quantitative differences between the simulations and the experiments are inevitable although the qualitative behaviour is expected to be captured. For instance, the flow velocities in the simulations of stable skyrmions can be much higher than those in the experiments, as shear flows in 2D are rather weak by comparison to 3D where these strong flows destroy the skyrmions.

The parameters used in the simulations are provided in Table S1 (in the supplemental material) both in simulation and physical units. We assume the material parameters of 5CB at room temperature~\cite{Ackerman2017, ANDERSON201515, yeh2009optics}. Some of these were converted from the parameters of the Ericksen-Leslie theory to the Beris-Edwards one using the expressions given in Ref.~\cite{marenduzzo07steady} and the equilibrium nematic order parameter $S_N=0.65$ (corresponding to the selected values of $a$, $b$ and $c$). For this set of parameters, corresponding to the experimental parameters of 5CB, the simulations remain stable at the cost of choosing a very small time step. The simulation domain is $L_X \times L_Y = 112 \times 56$ for the simulations of a single skyrmion and $L_X \times L_Y = 200 \times 200$ for the simulations of 30 skyrmions, with periodic boundary conditions in both directions. 

The initial configuration of each skyrmion is set as in Ref.~\cite{Coelho_2021} using an Ansatz for the director field: 
\begin{eqnarray}
 &&n_x = \sin(\tilde{a}) \sin(m\tilde{b} + g)\nonumber\\
 &&n_y = \sin(\tilde{a}) \cos(m\tilde{b} + g)\nonumber\\
 &&n_z = -\cos(\tilde{a}),
\end{eqnarray}
where
\begin{eqnarray}
 &&\tilde a =  \frac{\pi}{2}\left[1 - \tanh\left(\frac{B}{2}(\rho-R)\right)\right]\\
 && \tilde b = \tan^{-1}\left(\frac{x-C_x}{y-C_y}\right)\\
 && \rho = \sqrt{(x-C_x)^2+(y-C_y)^2} .
\end{eqnarray}
The parameter $R$ controls the size of the skyrmion, $B$ controls the width of the interface that separates the inner and the outer regions, $m$ is the winding number of the skyrmion, $g$ controls the direction of the skyrmion, $r$ measures the distance from the skyrmion centre and $\tilde b$ is the polar angle. We set the values of these parameters (in simulation units): $m=1$, $g = \pi/2$, $R=0.7p$, $B=0.5$, $C_x=L_X/2$, $C_y=L_Y/2$. The velocity field is set to zero while the Ansatz is relaxed until it reaches the steady state.

\subsection{Frank-Oseen free energy for quasi-static analysis}

We utilize the Frank-Oseen free energy to study the 3D director structures of stretched torons and CF-2s. For chiral nematic LCs, the free energy can be expressed as

\begin{align}
     F_{FO}= &\int d^3r \biggl (\frac{K_{11}}{2}(\nabla\cdot\nvec)^2+\frac{K_{22}}{2}(\nvec\cdot\nabla\times\nvec+\frac{2\pi}{p})^2+ \nonumber \\
    & \frac{K_{33}}{2}(\nvec\times\nabla\times\nvec)^2  - \frac{\varepsilon_0\Delta\varepsilon}{2} (\Evec \cdot \nvec)^2 \biggr),
    \label{eq:frank_oseen}
\end{align}
where the Frank elastic constant $K_{11}$, $K_{22}$ and $K_{33}$ determine the energy cost of splay, twist and bend deformations, respectively, and $p$ is the cholesteric pitch. The electric field $\Evec$ is along $z-$axis ($\norm{\Evec} = U/d$, with $U$ being the voltage across the cell and $d$ the cell thickness), $\varepsilon_0$ is the vacuum permittivity, and $\Delta\varepsilon$ is the dielectric anisotropy of the LC. Torons and CF-2s emerge as local or global minima of $F_{FO}$, and a relaxation routine based on the variational method is used to identify energy-minimizing configurations. For elongated torons of elongated length, we prepare initial structures by stretching the point defects of the original toron and letting the whole structure to relax. The 3D computations are performed on 40×100×40 square grids with periodic boundary conditions in the $x-$ and $y-$directions and with fixed boundary conditions at the bounding surfaces along the $z-$direction. For all calculations, the following values of the model parameters are used: $d/p = 1.5$, $K_{11}=6.4\times10^{-12} N, K_{22}=3\times10^{-12} N, K_{33}=10\time10^{-12} N, U = 0.85 V$ and $\Delta\varepsilon=13.8$. 

\subsection{Simulated polarized optical microscopy images}

The POM image is simulated by the Jones-matrix method, using the energy-minimizing configurations of $\nvec(\rvec)$ for the periodic CF-2s. The cell is split into 40 thin sublayers along the $z$ direction, then we calculate the Jones matrix for each pixel in each sublayer by identifying the local optical axis and ordinary and extraordinary phase retardation. The phase retardation arises from the optical anisotropy of LC ($n_o=1.58$ and $n_e=1.77$ for 5CB), where the optical axis is aligned with the local molecular direction. The Jones matrix for the whole LC cell is obtained by multiplying all Jones matrices corresponding to each sublayer. We obtain the single-wavelength POM by the respective component of the product of the Jones matrix and the incident polarization. To properly reproduce the experiment POMs, we produced images separately for three different wavelengths spanning the entire visible spectrum (450, 550, and 650$nm$) and then superimposed them, according to light source intensities at the corresponding wavelengths.

\subsection{Sample preparation and experimental methods}

Chiral LCs are prepared by mixing 4-Cyano-4'-pentylbiphenyl (5CB, EM Chemicals) with a left-handed chiral additive, cholesterol pelargonate (Sigma-Aldrich). To define the pitch ($p$) of the ensuing chiral LCs, the weight fraction of the added chiral dopant is
calculated using $C_{dopant}=1/(h_{htp})$, where the helical twisting power $h_{htp}=6.25 \mu m^{-1}$ for the cholesterol pelargonate additive. 

The sample cells are assembled from indium-tin-oxide (ITO)-coated glass slides treated with polyimide SE5661 (Nissan Chemicals) to obtain strong perpendicular (homeotropic) boundary conditions. The polyimide is applied to the surfaces by spin-coating at $2700 rpm$ for $30 s$ followed by baking (5 min at 90 $^\circ$C and then 1 h at 180$^\circ$C).

The LC cell gap thickness is defined by silica spheres as spacers between two surfaces to be 7 $\mu$m with the cell gap to pitch ratio $d/p = 1.25$. The lateral sides of the LC cell are sealed by epoxy except for the two entrances. We use a homemade connection obtained by 3D-printing with the Formlabs Form 2 resin 3D printer (purchased from Formlab) to connect one entrance and tube, the other end of the tube being connected to a syringe. Then by pushing or pulling the plunger, the LC inside the cell channel is driven forwards or backwards.  Metal wires were attached to ITO and connected to an external voltage supply (GFG-8216A, GW Instek) for electric control. 

\textcolor{black}{We utilize a ytterbium-doped fibre laser (YLR-10-1064, IPG Photonics, operating at $1064 nm$) to generate torons. The torons are three dimensional topological solitons where the skyrmion tube is embedded between the two substrates along the perpendicular axis, terminated by two point defects. We use a sinusoidal electric field with the frequency of $1000Hz$. Initially, we adjust the voltage across the cell to be around $U\approx 3.5 V$ where the LC is uniform and unwound. Then we melt the LC locally using a laser power around $30mW$ and switch off the laser tweezers, with torons being spontaneously generated after the LC quenching back. As we drive the LC flow forwards or backwards, the torons follow the flow.}

Polarizing optical microscopy images are obtained with a multi-modal imaging setup built around an IX-81 Olympus inverted microscope and charge-coupled device cameras (Grasshopper, Point Grey Research). Olympus objectives 20x and 10x with numerical aperture NA=0.4 and 0.1 are used. The speed, length, and trajectories of the torons are analysed by using ImageJ (freeware from NIH).

\subsection{Three-dimensional nonlinear optical imaging}

3D nonlinear optical imaging of periodic CF-2 structures is performed using the three-photon excitation fluorescence polarizing microscopy (3PEF-PM) setup built around the IX-81 Olympus inverted optical microscope integrated with the ytterbium-doped fibre laser \cite{Lee:10}. We use a Ti-Sapphire oscillator (Chameleon Ultra II; Coherent) operating at 870$nm$ with 140$fs$ pulses at an 80 MHz repetition rate as the source of the laser excitation light. An oil-immersion 40× objective (NA = 0.75) is used to collect the fluorescence signal. The fluorescence signal is detected by a photomultiplier tube (H5784-20, Hamamatsu) after a 417/60$nm$ bandpass filter. The 3PEF-PM imaging involves a third-order nonlinear process, during which LC molecules are excited via the three-photon absorption process and the signal intensity scales as $\propto \cos^6(\beta)$, where $\beta$ is the angle between the polarization of the excitation light and the long axis (transition dipole) of the LC molecule.

\section*{Acknowledgements}

We acknowledge financial support from the Portuguese Foundation for Science and Technology (FCT) under the contracts: PTDC/FISMAC/5689/2020, EXPL/FIS-MAC/0406/2021, UIDB/00618/2020 and UIDP/00618/2020.

\bibliography{ref}

\end{document}